\documentclass[letterpaper,
               nospread,
               biblatex,
              ]{jacow}
\makeatletter%
	\ifboolexpr{bool{xetex}}
	 {\renewcommand{\Gin@extensions}{.pdf,%
	                    .pdf,.jpg,.bmp,.pict,.tif,.psd,.mac,.sga,.tga,.gif,%
	                    .eps,.ps,%
	                    }}{}
\makeatother

\ifboolexpr{bool{xetex} or bool{luatex}} % test for XeTeX/LuaTeX
 {}                                      % input encoding is utf8 by default
 {\usepackage[utf8]{inputenc}}           % switch to utf8

\usepackage[USenglish]{babel}

\usepackage[final]{pdfpages}
\usepackage{multirow}
\usepackage{ragged2e}
\usepackage{subcaption}

\usepackage{todonotes}
\presetkeys{todonotes}{inline}{}
\ifboolexpr{bool{jacowbiblatex}}%
 {%
  \addbibresource{FRXD3.bib}
 }{}
\begin{document}

\title{Measurements of the five-dimensional phase space\\ distribution of an intense ion beam\thanks{This manuscript has been authored by UT- Battelle, LLC under Contract No. DE-AC05-00OR22725 with the U.S. Department of Energy. This research used resources at the Spallation Neutron Source, a DOE Office of Science User Facility operated by the Oak Ridge National Laboratory.}}

\author{A. Hoover\thanks{hooveram@ornl.gov}, K. Ruisard, A. Alexandrov, A. Zhukov, S. Cousineau, ORNL, Oak Ridge, TN, USA}
	
\maketitle

\begin{abstract}
No simulation of intense beam transport has accurately reproduced measurements at the level of beam halo. One potential explanation of this discrepancy is a lack of knowledge of the initial distribution of particles in six-dimensional (6D)~phase space. A direct 6D measurement of an ion beam was recently performed at the Spallation Neutron Source (SNS) Beam Test Facility (BTF), revealing nonlinear transverse-longitudinal correlations in the beam core that affect downstream evolution. Unfortunately, direct 6D~measurements are limited in resolution and dynamic range; here, we discuss the use of three slits and one screen to measure a 5D projection of the 6D phase space distribution, overcoming these limitations at the cost of one dimension. We examine the measured 5D distribution before and after transport through the BTF and compare to particle-in-cell simulations. We also discuss the possibility of reconstructing the 6D distribution from 5D and 4D projections.
\end{abstract}

\section{Introduction}

The phase space distributions of intense beams tend to develop halo during low-energy transport \cite{Batygin2021}. The existence of beam halo at higher energies leads to uncontrolled losses that limit the performance of high-power accelerators \cite{Henderson2014}. Attempts to reproduce measurements at the halo level in computer simulation have been unsuccessful \cite{Allen2002, Qiang2002, Groening2008}, likely due to an inaccurate initial distribution of particles in 6D~phase space.

The Spallation Neutron Source (SNS) Beam Test Facility (BTF) is poised to resolve these discrepancies. The BTF is a replica of the front-end of the SNS — the H$^-$ ion source, 65\,keV low-energy beam transport (LEBT), 402.5 MHz radio-frequency quadrupole (RFQ), and 2.5 MeV medium energy beam transport (MEBT) — followed by a 9.5-cell FODO transport line \cite{Zhang2020}. There are two emittance measurement stations: one 1.3 meters downstream of the RFQ and one after the FODO line. The two-dimensional (2D) horizontal and vertical phase space distributions can be measured at either station with a dynamic range above $10^6$ \cite{Aleksandrov2021}, large enough to image the beam halo \cite{Aleksandrov2020}. The 6D phase space distribution can be measured at the first station \cite{Cathey2018}. 

One goal of the research program at the BTF is to reproduce the measured halo at the second emittance station in simulation. Unfortunately, it is not currently possible to generate a realistic initial bunch due to the limited resolution (10~points per dimension) and dynamic range ($10^1$) of 6D measurements.\footnote{A 6D measurement with these parameters took over thirty-two hours in Ref. \cite{Cathey2018} but has been reduced to twenty-four hours with improved data acquisition software.} Improvements to the scan efficiency are under development; in this paper, we consider the use of three slits and one screen to measure a 5D projection of the 6D phase space distribution, overcoming these limitations at the cost of one dimension. We examine the measured 5D~distribution at both emittance stations and discuss how the measurements might contribute to the goal of halo prediction at the BTF.

\section{Measurements of the initial distribution in the BTF}

The measurement apparatus is described in \cite{Ruisard2022-NAPAC}; it consists of two vertical slits to select the horizontal position $x$ and momentum $x'$, a vertical slit to select the vertical position $y$, and a scintillating screen after a dipole bend. The vertical momentum $y'$ is linearly related to the vertical position on the screen; the energy deviation $w$ is a function of the horizontal position on the screen and the $x$ and $x'$ selected by the upstream slits. The scan consists of a series of "sweeps" where the vertical slits are held fixed while the horizontal slit moves at a fixed speed, collecting the image on the screen on every beam pulse at a 5 Hz repetition rate. A rectilinear scan pattern is employed with a linear correlation between $x$ and $x'$ to align the grid with the $x$-$x'$ distribution. The data is linearly interpolated onto a regular grid after transforming to phase space coordinates.

\subsection{Revisiting Hidden Features in the Beam Core}

We first examine a measured 5D distribution at the first emittance station. 57,766 images were collected over 4.9~hours as the slits traversed a $32 \times 32 \times 32$ grid. The beam current during the measurement was stable at 30.6 mA. The full projections of this data set are shown in \cite{Ruisard2022-NAPAC}; the density profiles vary smoothly in all projections and there are no visible inter-plane correlations.

\begin{figure}[!t]
    \centering
    \begin{subfigure}{\columnwidth}
        \includegraphics[width=\textwidth]{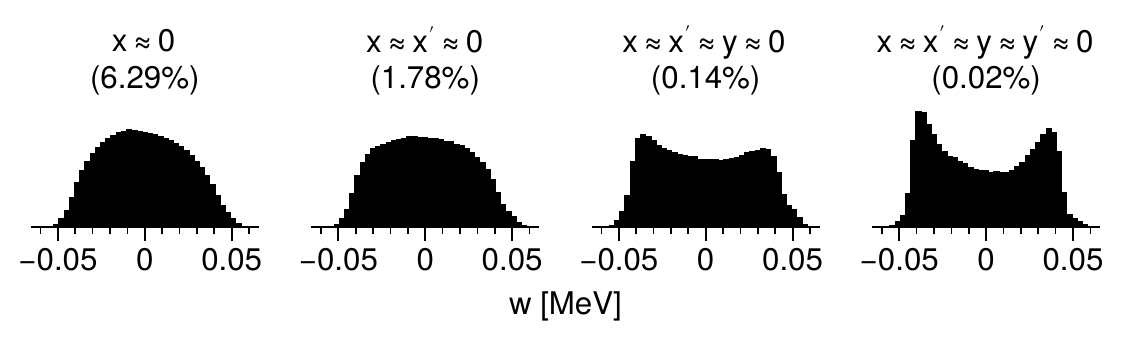}
        \caption{}
        \label{fig:hollow_a}
    \end{subfigure}
    \begin{subfigure}{\columnwidth}
        \includegraphics[width=\textwidth]{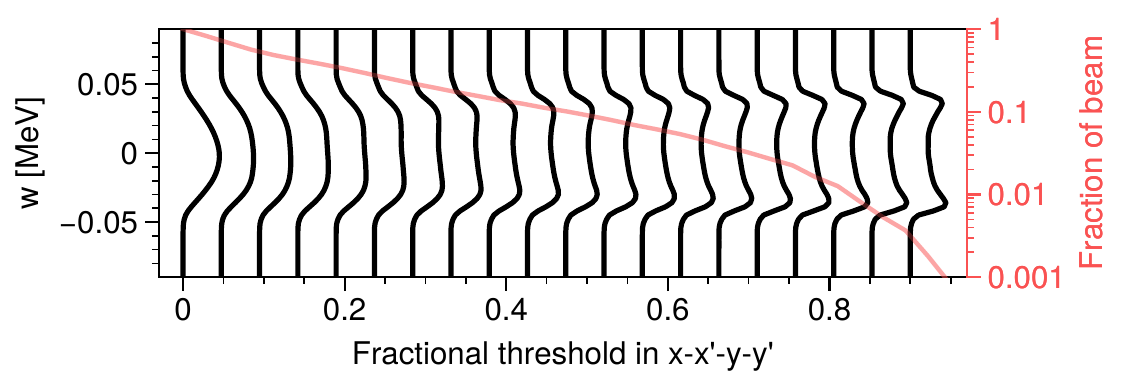}
        \caption{}
        \label{fig:hollow_b}
    \end{subfigure}
    \caption{Hollow energy distribution in the beam core measured at the first emittance station in the BTF. (a) Energy distribution of particles within progressively smaller slices in transverse phase space. The approximate fraction of beam particles selected are shown in parenthesis. (b) Energy distribution (black) and fraction of beam (red) within a shrinking density contour in transverse phase space.}
    \label{fig:hollow}
\end{figure}

Yet previous high-dimensional measurements of the beam exiting the RFQ have revealed nonlinear transverse-longitudinal correlations in the beam core: the energy and phase distribution of particles at the center of transverse phase space ($x \approx x' \approx y \approx y' \approx 0$) is hollow and bimodal while the full projection is unimodal \cite{Cathey2018}. This is a space-charge-driven effect \cite{Ruisard2021-hollow} and has a clear dependence on the beam intensity. The relationship between the energy and transverse coordinates has been examined by slicing the distribution in the transverse plane before projecting the distribution onto the energy axis; i.e., by inserting slits upstream of the longitudinal emittance measurement diagnostic. The five-dimensional nature of the correlation is made clear by observing the energy profile as each dimension is sliced. We repeat this for our 5D measurement in Fig.~\ref{fig:hollow}(a).
\begin{figure*}[!t]
    \centering
    \begin{subfigure}{0.48\textwidth}
        \includegraphics[width=\textwidth]{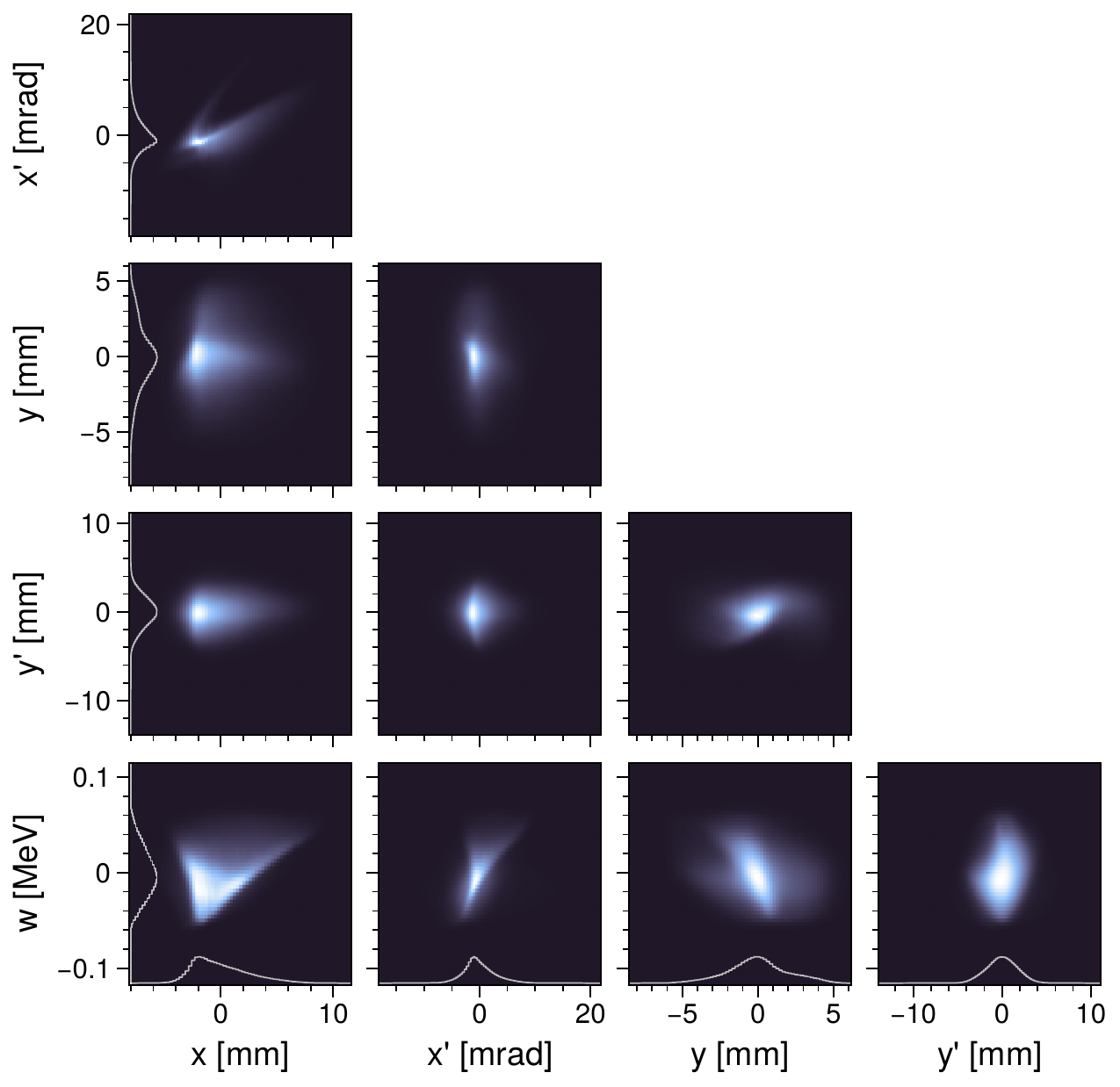}
        \caption{}
        \label{fig:VS34_a}
    \end{subfigure}
    \hfill
    \hspace{0.1cm}
    \hfill
    \begin{subfigure}{0.48\textwidth}
        \includegraphics[width=\textwidth]{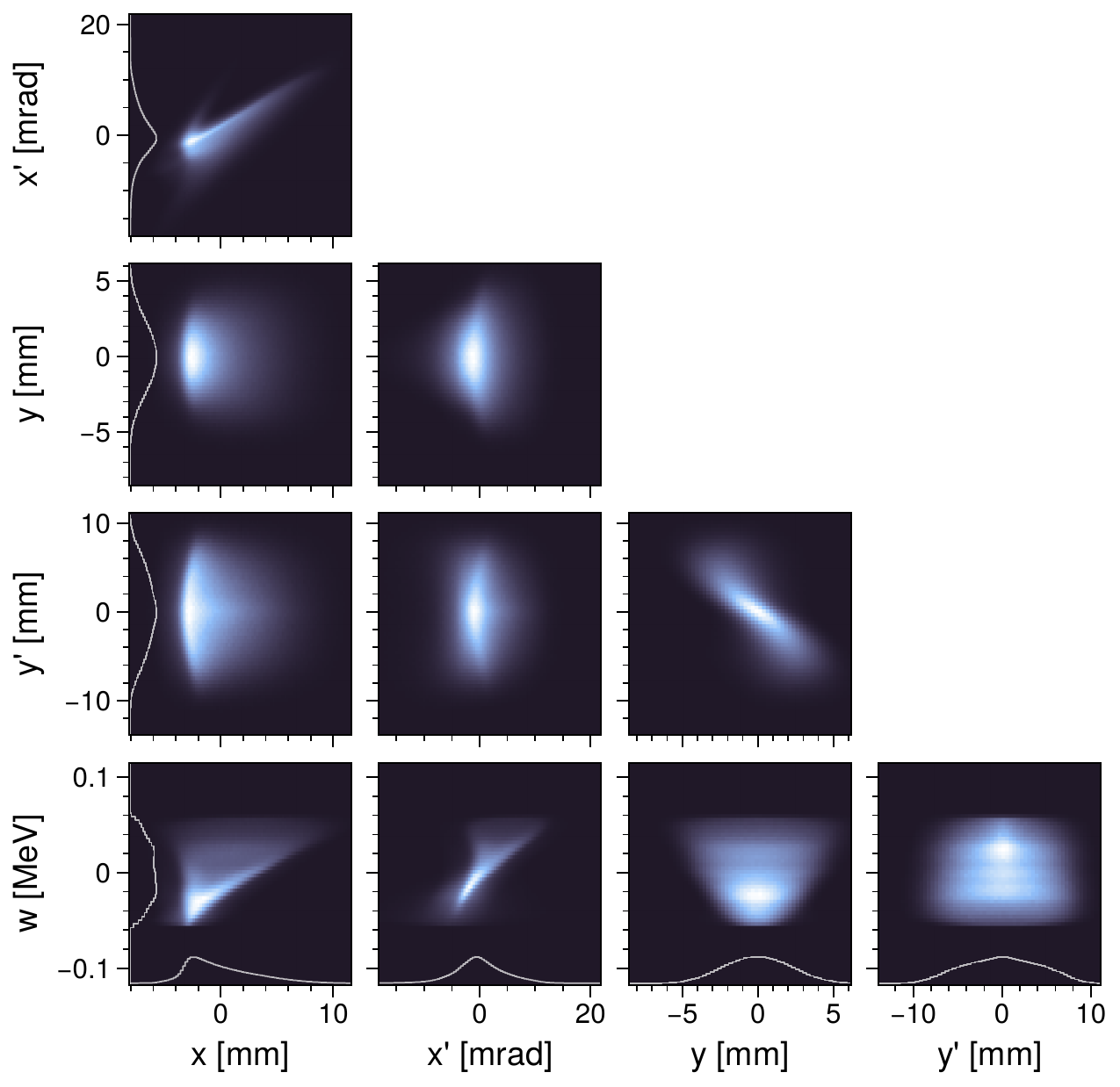}
        \caption{}
        \label{fig:VS34_b}
    \end{subfigure}
    \vfill
    % \vspace*{0.2cm}
    \vfill
    \begin{subfigure}{\textwidth}
        \includegraphics[width=\textwidth]{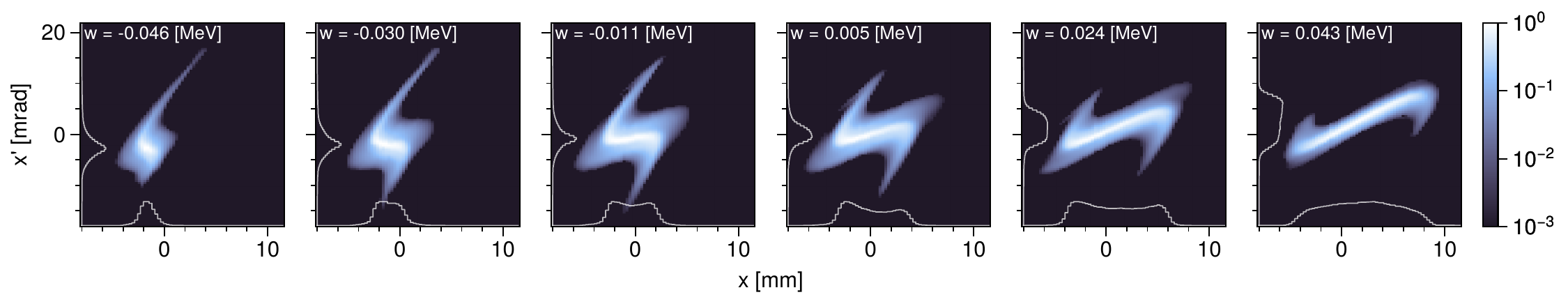}
        \caption{}
        \label{fig:VS34_c}
    \end{subfigure}
    \vfill
    % \vspace*{0.2cm}
    \vfill
    \begin{subfigure}{\textwidth}
        \includegraphics[width=\textwidth]{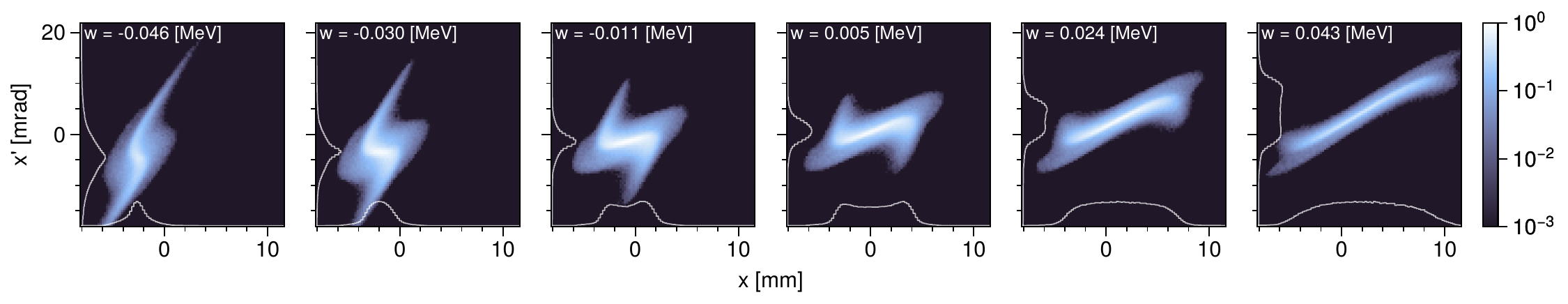}
        \caption{}
        \label{fig:VS34_d}
    \end{subfigure}
    \caption{5D measurement-simulation comparison after transport through the BTF. The measured (a) and simulated (b) projections are shown in linear color scale. The measured (c) and simulated (d)~$x$-$x'$ projections within $\approx$ 2 keV energy slices are shown in logarithmic color scale. (Note that the color scales are not normalized across subplots.)}
    \label{fig:VS34}
\end{figure*}

Slicing the array in this way selects only a small fraction of the beam particles — vanishingly small when three or four dimensions are sliced — and does not characterize the extent of the hollow-energy core in the full transverse phase space. This information is encoded in the 5D measurement: Fig.~\ref{fig:hollow}(b) plots the energy distribution within a shrinking volume of transverse phase space. For each threshold $t$, we project the 5D array $f_{5D}$ along the $w$ axis to obtain $f_{4D}$, normalize the elements of $f_{4D}$ to the range $[0, 1]$, locate the indices $\left\{i, j, k, l\right\}$ such that $f_{4D}[i, j, k, l] < t$, and mask these indices in $f_{5D}$ before projecting $f_{5D}$ onto the $w$ axis. The energy distribution transitions smoothly from unimodal to bimodal as $t$ increases; two peaks begin to emerge at $t \approx 0.28$, a volume containing 22\% of the beam particles.

\subsection{Reconstructing the 6D Distribution from 5D and 4D Projections}

The effect of hidden high-dimensional features on halo formation and long-term beam dynamics is not yet known due to the lack of a realistic 6D simulation bunch.\footnote{
   Bunches produced by LEBT/RFQ simulations have qualitatively similar features to those seen in high-dimensional measurements, albeit large errors in the root-mean-square emittances. The simulated transport of these bunches through the SNS linac indicates that the correlations affect the root-mean-square beam sizes over time \cite{Ruisard2021-IPAC}.
} The present resolution and dynamic range of direct 6D measurements are too low to capture these features. It is therefore prudent to determine whether the 6D distribution can be recovered from the high-resolution, relatively high-dynamic-range 5D measurements described herein. The problem can be formulated as follows, letting $z$ and $z'$ be the longitudinal phase space coordinates: construct $f(x, x', y, y', z, z')$ consistent with the measured $f(x, x', y, y', z')$ and $f(x, x', z, z')$, where the latter is measured using two vertical slits and a BSM \cite{Ruisard2020}. 

We suggest several approaches here. A simple but promising approach is to generate a 5D particle distribution, then add a $z$-$z'$ correlation ``by hand", assuming $z \approx az' + \Delta$ where $a$ is a constant and $\Delta$ is a small random variable. This reflects the longitudinal phase space at the measurement location in the BTF and will be our first attempt. More interesting is the case of an arbitrary $z$ distribution. An almost identical problem has been encountered in astrophysics: the Gaia satellite has measured the phase space coordinates of many stars in the Milky Way galaxy, but a large fraction of the measurements contain only five of the six coordinates; neural networks have been employed to infer the missing velocity coordinate \cite{Dropulic2021}. Another approach is to treat the problem as a tomographic image reconstruction; for example, the Maximum Entropy (MENT) algorithm could be used, which selects the maximum entropy image among those consistent with the measured projections \cite{Skilling1984}. It is unknown if MENT (or other image reconstruction algorithms) can easily scale from two to six dimensions.

\section{5D simulation benchmark}

While the 6D measurement can only be performed at the first emittance station, the 5D measurement can be performed at either emittance station. This provides a unique opportunity for a 5D simulation benchmark. We performed a scan at the second emittance station, collecting 220,783~images over 13.8 hours. The beam current was stable during the scan at 25.8 mA. This produced an interpolated phase space density array with dimensions 71 $\times$ 104 $\times$ 71 $\times$ 110 $\times$ 92.\footnote{
    The measurement grid is sheared along the $x'$ axis, boosting the resolution along the second axis of the array. All images were downscaled by a factor of three using local averaging.
}

Figures \ref{fig:VS34}(a) and \ref{fig:VS34}(b) compare the 1D and 2D projections of the measured 5D distribution with a particle-in-cell simulation. The simulation was performed in the PyORBIT code with a 3D fast Fourier transform space charge solver on a 64 $\times$ 64 $\times$ 64 mesh with 10,000,000 macro-particles. The initial simulation bunch was generated by independently sampling 2D measurements such that $f(x,x',y,y',z,z') = f(x,x')f(y,y')f(z,z')$.\footnote{
    The 2D measurements were taken at a beam current of 29 mA, slightly higher than in the 5D measurement. We used a 29 mA beam current in this simulation.
} The reasonable agreement in $x$-$x'$ was expected from previous work \cite{Ruisard2021-IPAC}. The discrepancy in $y$-$y'$ was also expected; it may be due to unidentified quadrupole misalignments or an offset of the beam centroid. 

Strong correlations in $x$-$x'$-$w$ subspace are evident in the 2D projections. Figures \ref{fig:VS34}(c) and \ref{fig:VS34}(d) examine this subspace in more detail, displaying the $x$-$x'$ distribution within $\approx$ 2\,keV energy slices. Although the simulation-measurement agreement is not perfect, the elongation and rotation of the $x$-$x'$ projection as a function of energy is reproduced. A detailed investigation of the measured 5D phase space distribution is reserved for a future publication.

\section{Conclusion}

The Spallation Neutron Source (SNS) Beam Test Facility (BTF) has the ability to measure the full 6D phase space distribution of an intense ion beam. Such a measurement could lead to the accurate prediction of beam halo formation; however, the resolution and dynamic range of 6D measurements are currently too low to generate a realistic simulation bunch. 5D phase space measurements offer significant improvements in dynamic range and resolution. The 5D phase space distribution was measured at the first emittance station in the BTF. Hidden high-dimensional features in the beam core, previously measured only in small slices of the transverse phase space, were revisited in the 5D distribution. Several strategies to reconstruct the 6D distribution from the 5D measurement were suggested. Finally, a particle-in-cell simulation was benchmarked against a 5D measurement at the end of the BTF. Reasonable agreement was found in $x$-$x'$-$w$ space; disagreement remains in $y$-$y'$ space.

\printbibliography

\end{document}